# Prediction of Separation Induced Transition on Thick Airfoil Using Non-linear URANS Based Turbulence Model

## Alok Mishra, Gaurav Kumar, and Ashoke De[*]

*Department of Aerospace Engineering, Indian Institute of Technology Kanpur, Kanpur, India, 208016*

**Abstract**

Most of the turbulence models in practice are based on the assumption of a linear relation between Reynolds stresses and mean flow strain rates which generally provides a good approximation in case of attached and fully turbulent flows. However, this is seldom the need in most of the engineering problems; the majority of the engineering problems observe flow separation or flow transition. Recent developments in non-linear turbulence models have proven significant improvement in prediction of separated flow due to better resolution of anisotropy in modeled Reynolds stress. The domain of application of this improved RANS model can be extended to flow transitions as well, where the resolution of anisotropy in Reynolds stress is required. For a validation of such kind, a two-dimensional numerical study has been carried out over NACA 0021 with k-ω SST model with non-linear correction at Re = 120,000 for various angles of attack which experiences the formation of a Laminar Separation Bubble (LSB). A correct prediction of LSB requires an accurate resolution of anisotropy in Reynolds stresses. For comparison with other linear models, the simulations are also performed with k-$k_l$-ω (a 3-equation linear transition model), k-ω SST (a 2-equation linear model) and Spalart-Allmaras (a 1-equation model). The performance of these models is assessed through aerodynamic lift, drag, pressure and friction coefficients. It is found that the non-linear k-ω SST and k-$k_l$-ω transition model provide comparable quality of prediction in lift and drag coefficients (in spite of the fact that non-linear k-ω SST involves solving less number of transport equation than the transition model) as observed in the experiments whereas k-ω SST and SA models under predict the drag coefficient value at low angle of attack due to inability to capture the separation induced transition. It is also observed that the location of laminar separation bubble is captured accurately when non-linear or transition model is used as opposed to the SA or linear SST models, which lack in the ability to predict the same.

## 1. Introduction

In recent times, the interest in flows with a low Reynolds number has increased due to renewed interest in micro air vehicle, high altitude sailplane, wind turbine and turbomachinery blades. The transition, a salient phenomenon, occurs in this range of Reynolds number. Boundary layer over an airfoil at low Reynolds number may still be laminar, but due to poor resistance to the adverse pressure gradient, flow separation from the airfoil takes place and transition occurs in the mid-air which subsequently leads to turbulent flow. If the turbulent shear layer has sufficient energy, it may re-attach with the airfoil as a turbulent shear layer and the air entrapped between the separation and re-attachment points are called laminar separation bubble. Laminar separation bubble decreases the performance of airfoil and also affects its stall characteristics [1-6]. Hence, accurate prediction of the transition is very crucial in improving the performance of airfoil at low Reynolds number range.

Briley [7] was one of the first who numerically studied the laminar separation bubble using the Navier-Stokes equations for incompressible flow over a flat plate. The flow in the vicinity of separation bubble was calculated using the Navier-Stokes equations and the flow over the rest of domain was investigated through boundary layer theory and inviscid flow analysis. The numerical results were compared with the solution of boundary layer equation and it was found that the boundary layer theory is not reliable for the prediction of laminar separation bubble at low Reynolds number range. It was also suggested that the similar study can be applied on airfoil to investigate transition.

Briley and McDonald [8] developed a numerical method for comprehensive computations of thin incompressible separation bubbles on smooth surfaces with less computation cost. The method also provides a realistic detailed description of the location of transition and mean flow behaviour during and after the transition, including the sub-layer region. Numerical simulation was found to be in good agreement with experimental data for transitional separation bubbles on a NACA 66₃-018 airfoil at zero angle of incidence with chordal Reynolds numbers of $1.7 \times 10^6$ and $2 \times 10^6$. Crimi and Reeves [9] analyzed the small separation bubble near the leading edge and developed correlation of local shear stress layers to determine the onset of transition in laminar separation. Roberts [10] developed a modified semi-empirical theory to predict the growth and bursting of laminar separation bubble on NACA 66₃-018 airfoil at various angles of attack and results showed good agreement with the experimental data. The theory is also helpful to evaluate approximate bursting Reynolds number, bubble length and abrupt stall.

Krumbein[11-13] coupled a Reynolds-averaged Navier-Stokes (RANS) solver, a laminar boundary layer code and two e^N-database methods for the prediction of transition. The numerical studies have been performed for the developed RANS on high-lift multi-element airfoil configurations and 3D finite wing. The results were validated against experimental data and found in good agreement. Langtry and Menter [14] developed a correlation based model which is based on the local variable. The transition model was tested for various cases to assess the transition capability. It was found that the model is fully CFD-compatible and does not have any adverse effect on the convergence of solver. Rogowski et al. [15] utilized correlation based transition model and RNG model to simulate NACA0018 airfoil and Darrieus-type wind turbine. It was found that the transition model provides close results to experimental data as compared to RNG k-ε model. Catalano and Tognaccini [16] proposed a modification in k-ω SST in order to improve simulation at low Reynolds number flows. The modified RANS model was applied to simulate flow over Selig-Donovan 7003 airfoil at Reynolds number 6×10⁴. For prediction of laminar separation bubble, the simulated RANS model exhibited good agreement with LES data. Walter and Cokljat [17-18] modified two equations linear eddy viscosity turbulence model in order to develop new transition model for bypass and natural transition prediction. A new transport equation was added to represent the growth of non-turbulent, stream wise fluctuation in the pre-transitional boundary layer. The RANS based 3-equation transition model was validated against the experimental data and results show that the model exhibited good prediction capability of transition.

Choudhry et al. [19] investigated the flow around NACA 0021 airfoil with the correlation-based γ-Re$_θ$ model and the laminar-kinetic-energy-based k-$k_l$-ω model. Overall performance of k-$k_l$-ω model is superior over the γ-Re$_θ$ model. The γ-Re$_θ$ model predicts a sharp increase in turbulence level which causes fast transition over the airfoil and early turbulent reattachment. The length of separation bubble reduces with increase in Reynolds number. The effect of increasing Reynolds number on laminar separation bubble is similar to the increase in turbulent intensity.

Even though several numerical studies have been done with RANS models in prediction of laminar separation bubble, there exists a gap in the literature for analysis of the behavior of transition phenomenon over NACA 0021 airfoil. It is quite obvious that the transition model is usually invoked to predict laminar separation bubble and/or transition behavior, but the addition of non-linear corrections to k-ω SST model introduces the capability to predict transition over linear k-ω SST. A paucity of study available in the open literature regarding the capability of non-linear correction in k-ω SST model in predicting the transition behavior has motivated the authors to carry out the current analysis. This includes a comparative study of the behavior of k-ω SST, non-linear k-ω SST, k-$k_l$-ω and SA models for NACA 0021 airfoil operating at Re = 120,000 and also to verify the prediction capability of non-linear k-ω SST model for transitional flows.

## 2. Mathematical Model

### 2.1 Flow field equation

The governing equations for the conservation of mass and momentum are given by

$$\frac{\partial \bar{U}_i}{\partial x_i} = 0 \tag{1}$$

$$\frac{\partial \bar{U}_i}{\partial t} + \frac{\partial \bar{U}_i \bar{U}_j}{\partial x_j} = -\frac{1}{\rho}\frac{\partial \bar{p}}{\partial x_i} + \nu \frac{\partial^2 \bar{U}_i}{\partial x_i \partial x_j} + \frac{\partial \tau_{ij}}{\partial x_j} \tag{2}$$

Here ‾ is used to represent the Reynolds averaged variables. $\bar{U}_i$ is the fluid velocity, $\bar{p}$ is the fluid pressure, $\nu$ is the kinematic viscosity and $\tau_{ij}$ is the Reynolds stress tensor. This set of equations is closed by modeling of Reynolds stress term in the momentum equation by different turbulence models as explained in the following subsection.

### 2.2 Spalart-Allmaras (SA) model

The Spalart-Allmaras model is a one-equation mixing-length model, developed for external flows by Spalart and Allmaras [20] and commonly used for external aerodynamic calculations. The Boussinesq eddy viscosity assumption is used to calculate Reynolds stresses. The model solves a transport equation for the turbulent viscosity ($\bar{\nu}$). A detailed description of one equation mixing-length model can be found in [20].

### 2.3 Shear Stress Transport Model (k-ω SST)

The k-ω SST model [21] uses the k-ω equations in the regions of the boundary layer and switches to the k-ε model elsewhere and both models are combined with the help of blending function F$_1$. The equations for turbulent kinetic energy (k) and specific turbulent dissipation rate (ω) are given below

$$\frac{\partial k}{\partial t} + \overline{u}_j \frac{\partial k}{\partial x_j} = P_k - \beta^* \omega k + \frac{\partial}{\partial x_j}\left[\left(\nu + \sigma_k \nu_t\right)\frac{\partial k}{\partial x_j}\right] \qquad (3)$$

$$\frac{\partial \omega}{\partial t} + \overline{u}_j \frac{\partial \omega}{\partial x_j} = \frac{\gamma}{\nu_t} P_k - \beta \omega^2 + \frac{\partial}{\partial x_j}\left[\left(\nu + \sigma_\omega \nu_t\right)\frac{\partial \omega}{\partial x_j}\right]$$
$$+ 2\left(1 - F_1\right)\sigma_{\omega 2} \frac{1}{\omega}\frac{\partial k}{\partial x_j}\frac{\partial \omega}{\partial x_j} \qquad (4)$$

where $P_k = \max\left(\tau_{ij}\,\partial \overline{u}_i/\partial x_j\,, 10\beta^*\omega k\right)$, is the kinetic energy production term, $\beta^* = 0.09$ is a model constant and the last term in the $\omega$ equation is the cross-diffusion term. $F_1$ is a blending function which has a value of one inside the boundary layer and zero outside. The limiting of the production term is an alternative to the use of damping function in the near-wall region. The turbulent stress tensor and viscosity are computed in this model as follows

$$\tau_{ij} = \frac{2}{3}k\delta_{ij} - 2\nu_t \overline{S}_{ij} \qquad (5)$$

More insight of the shear stress transport model is given in [21].

### 2.4 Non-linear Shear Stress Transport Model (NSST)

In non-linear k-$\omega$ SST model, a non-linear Reynolds stress component ($N_{ij}$) is added to Reynolds stress tensor. The expression for $\tau_{ij}$ is given in Eq. 6

$$\tau_{ij} = \frac{2}{3}k\delta_{ij} - 2\nu_t \overline{S}_{ij} + N_{ij} \qquad (6)$$

Here $N_{ij}$ is the non-linear part of the Reynolds stress tensor. It had been set to zero for the linear model. For the non-linear k-$\omega$ SST model (NSST), a modified form of the non-linear constitutive relation proposed by Abe et al. [22] is used. The non-linear term is defined as follows

$$N_{ij} = f_{NL}\frac{3\nu_t^2}{k}\left[\begin{array}{c} f_s\left(2\overline{S}_{ik}\overline{S}_{kj} - \frac{2}{3}\overline{S}_{nk}\overline{S}_{kn}\delta_{ij}\right) \\ -\overline{S}_{ij}\Omega_{kj} - \overline{S}_{ik}\Omega_{ki} \end{array}\right] + 2kd_{ij}^\omega \qquad (7)$$

$$f_{NL} = \frac{4}{3}C_D C_B \left(1 - f_\omega(26)\right) \qquad (8)$$

$$C_B = \frac{1}{1 + 22\!\big/\!3\left(C_D\nu_t\big/k\right)^2\Omega^2 + 2\!\big/\!3\left(C_D\nu_t\big/k\right)^2\left(\Omega^2 - S^2\right)f_B} \qquad (9)$$

$$f_\omega(\eta) = \exp\left(-\left(y^+\big/\eta\right)\right)$$

$$f_s = 1 - \frac{S^2\left(\Omega^2 - S^2\right)}{\left(\Omega^2 + S^2\right)}\left\{1 + C_{s2}C_D\left(\Omega - S\right)\frac{\nu_t}{k}\right\} \qquad (10)$$

where $\Omega = \sqrt{\Omega_{ij}\Omega_{ij}}$ is the characteristic rotation rate, $S = \sqrt{\bar{S}_{ij}\bar{S}_{ij}}$ is the characteristic strain rate, $f_B = 1 + C_\eta C_D \nu_t / \left( k(\Omega - S) \right)$, $y^+ = u_\tau d_y / \nu$ is a non-dimensional wall distance, $\eta$ is a parameter, and $C_D = 0.8$, $C_\eta = 100$, and $C_{s2} = 7$ are model constants. Finally, the expression for the term $d_{ij}^\omega$ can be written as

$$d_{ij}^\omega = -\alpha_\omega f_\omega (26) \frac{1}{2} \left( d_i d_j - \frac{\delta_{ij}}{3} d_k d_k \right)$$

$$+ f_\omega (26) \left( 1 - f_{r1}^2 \right) T_d^2 \left\{ -\frac{\beta_\omega C_\omega}{1 + C_\omega T_d^2 \sqrt{S^2 \Omega^2}} \left( \bar{S}_{ik}\Omega_{kj} - \Omega_{ik}\bar{S}_{kj} \right) \right\}$$

$$+ f_\omega (26) \left( 1 - f_{r1}^2 \right) T_d^2 \left\{ \frac{\gamma_\omega C_\omega}{1 + C_\omega T_d^2 S^2} \left( \bar{S}_{ik}\bar{S}_{kj} - \frac{\delta_{ij}}{3} S^2 \right) F_i \right\} \quad (11)$$

$$f_{r1} = \left( \Omega^2 - S^2 \right) \Big/ \left( \Omega^2 + S^2 \right) \quad (12)$$

$$T_d = \left\{ 1 - f_\omega (15) \right\} k/\varepsilon + f_\omega (15) \delta_\omega \sqrt{\nu/\varepsilon} \quad (13)$$

$$\alpha_\omega = 1, \beta_\omega = \frac{1}{4}, C_\omega = 0.5, \gamma_\omega = 1.5, and\ \delta_\omega = 1.0 \quad (14)$$

Where $d_i = \partial N_i / \partial x_j$, $N_i$ is the unit-normal, and $\varepsilon = \beta^* \omega k$ is the turbulence dissipation term. The model constants for the term $d_{ij}^\omega$ are given in Eq. (14). This non-linear correction to SST model is adopted from Abe et al. [22] and more details on the exact expression used in this study can be found in Kumar et al. [23,24].

## 2.5 Laminar kinetic energy model (k-$k_l$-ω)

The k-$k_l$-ω transition model is a linear eddy viscosity model, based on laminar and turbulent kinetic energy. In addition to the continuity and momentum equations, three transport equations are solved for turbulent kinetic energy ($k_T$), the laminar kinetic energy ($k_L$) and specific dissipation rate (ω). The omega is defined as $\omega = \varepsilon / k_T$ where ε is the isotropic dissipation. The three transport equations are given below in Eqs. 15-17.

$$\frac{\partial k_T}{\partial t} + \bar{u}_j \frac{\partial k_T}{\partial x_j} = P_{k_T} + R_{BP} + R_{NAT} - \omega k_T$$

$$- D_T + \frac{\partial}{\partial x_j} \left[ \left( \nu + \frac{\alpha_T}{\sigma_k} \right) \frac{\partial k_T}{\partial x_j} \right] \quad (15)$$

$$\frac{\partial k_L}{\partial t} + \bar{u}_j \frac{\partial k_L}{\partial x_j} = P_{k_L} - R_{BP} - R_{NAT}$$

$$- D_L + \frac{\partial}{\partial x_j} \left[ \nu \frac{\partial k_L}{\partial x_j} \right] \quad (16)$$

$$\frac{\partial \omega}{\partial t} + \bar{u}_j \frac{\partial \omega}{\partial x_j} = c_{\omega 1} \frac{\omega}{k_T} P_{k_T} + \left( \frac{C_{\omega R}}{f_\omega} - 1 \right)$$

$$\frac{\omega}{k_T} \left( R_{BP} + R_{NAT} \right) - C_{\omega 2} \omega^2$$

$$+ C_{\omega 3} f_\omega \alpha_T f_\omega^2 \frac{\sqrt{k_T}}{d^3} + \frac{\partial}{\partial x_j} \left[ \left( \nu + \frac{\alpha_t}{\sigma_\omega} \right) \frac{\partial \omega}{\partial x_j} \right] \quad (17)$$

The various terms in above equations represent production, dissipation and transport mechanism of kinetic energy and specific dissipation rate. More details about this transition model can be found in [18].

## 3. Numerical Details

A rectangular domain of size 30C×30C is utilized in present simulations which is shown in Figure 1. The origin of the coordinate system is located at the leading edge of the airfoil. The flow inlet and outlet boundaries are located at 10C upstream and 20C downstream, respectively, from the leading edge of the airfoil. The top and bottom boundaries are located at a distance of 15C from the origin. The lateral distances are adequate to ensure the stability of simulation and also to eliminate the far field boundary effects.

### 3.1 Grid Generation

Grids used in the current study are generated using commercial meshing software ANSYS® ICEM-CFD [25]. A total of three different grids were generated namely coarse, medium and fine grids. Figure 2 depicts the grid around airfoil, where the average value of $y^+$ is found to be less than 1 for all the grids.

### 3.2 Numerical Solver

All the simulations are performed using the open-source CFD toolbox OpenFOAM® [26]. Pressure-velocity coupling is achieved using the PISO algorithm available in the code. The convection term in the momentum and turbulent variable equations are discretized using a bounded second-order scheme. Time marching is performed using a second-order backward difference scheme. All other terms are discretized using central difference schemes. The algebraic equation for velocity and pressure are inverted using a coupled solver and Preconditioned conjugate-gradient method, respectively. For the turbulent variables, a Gauss-Seidel scheme is employed. The tolerance has been set to $10^{-6}$ for all the variables. Once a statistically steady state is achieved, time averaging is performed over 100 flow-through times ($C/U_\infty$) to compute the statistics.

### 3.3 Boundary conditions

Freesteam boundary condition is specified at the inlet for velocity and pressure; while at the outlet, the convective boundary condition is applied. Top and bottom boundaries are also considered as freestream boundary condition. No-slip boundary condition is enforced at the airfoil boundary surface and pressure is set to zero-gradient. As the value of $y^+$ lies in the viscous layer of the boundary layer, no wall-function is employed and the turbulence variables are directly integrated up to the wall.

For k-$k_l$-ω model, at solid boundaries, a no slip boundary condition enforces $k_L = k_T = 0$. For ω, a wall normal zero-gradient condition is used. The turbulence intensity measured near the leading edge in the experiments is 0.8% as provided by Hansen et al. [27] and same is used to calculate the kinetic energy. At inlets, sufficiently far from solid walls, the laminar kinetic energy associated with pre-transitional fluctuations is zero and the appropriate boundary condition is specified for other variables.

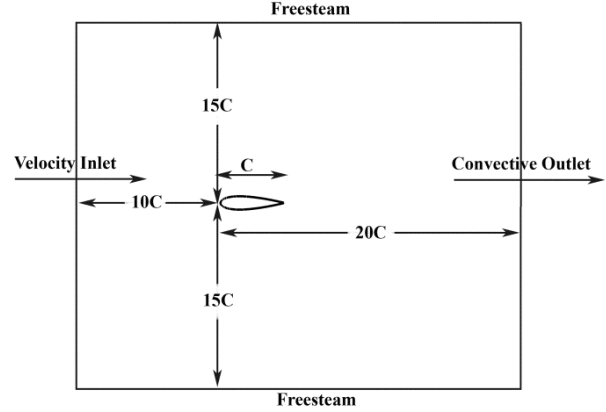

Fig. 1. Computational setup for the simulations

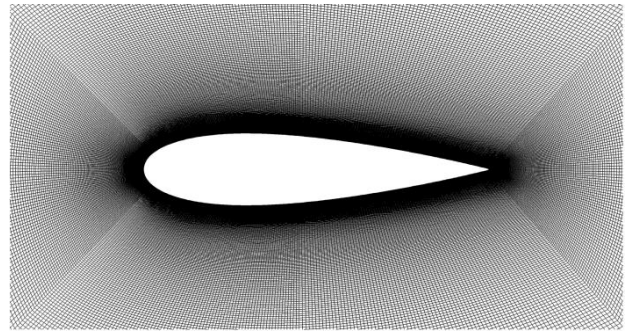

Fig. 2. Grid used for computations

For k-ω SST and NSST model, boundary conditions for turbulent kinetic energy and specific turbulence dissipation rate are same as the k-$k_l$-ω model. A wall function is applied for ω boundary condition on the airfoil which does nothing but applies a calculated value of omega at the wall using the expression $\omega_{wall} = 60\nu/\beta y^2$ where β = 0.09 and y is the distance of the first node from the wall.

For SA model, turbulent viscosity ($\tilde{v}$) is calculated as $\tilde{v} = v$ and specified on the inlet boundary. At the walls, this value is specified as zero.

### 3.4 Grid independence test

Grid comparison has been performed with three different grids at 0 and 10 degrees angle of attack (AOA) using k-$k_l$-ω model which is shown in Table 1. The k-$k_l$-ω model is well established model to predict transition and this is the reason behind selecting k-$k_l$-ω transition model for grid independence test. The grids are systematically refined by increasing the nodes on the airfoil surface. In this process, the $y^+$<1 is maintained for all the grids. The generated grids are tested for two angles of attack, i.e. 0 and 10 degrees, to check the dependency of the solution on the grids. The lift coefficient ($C_L$), Drag coefficient ($C_D$) and Pressure coefficient ($C_p$) are monitored and compared to different grids. The $C_p$ plots for 0 and 10 degrees AOA for k-$k_l$-ω are shown in Figure 3(a) and 3(b).

It can be observed from Table 1 that there is no significant difference, among the grid choices, for the global quantities such as $C_L$ and $C_D$. From Figure 1(a) and 1(b), effect of laminar separation bubble on the surface pressure coefficient is reflected in the predicted location and magnitude of pressure drop. This is due to the sensitivity of numerics with the grid resolution at a high angle of attack. In surface pressure coefficient plot for 0 degree AOA, this difference is negligible for all the grid resolutions.

However, for 10 degrees AOA, pressure drop prediction for laminar separation bubble is severely distorted for the coarse grid, whereas medium and fine grids have a very comparable prediction. Considering above details for the selection of optimum grid, the medium grid has been chosen and pursued for all the further studies.

Table1: Grids used for the simulations

| Grid | No. of cells | 0 deg AOA | | 10 deg AOA | |
|---|---|---|---|---|---|
| | | $C_L$ | $C_D$ | $C_L$ | $C_D$ |
| Coarse | 144,151 | 0.0062 | 0.021 | 0.885 | 0.053 |
| Medium | 188,461 | 0.0068 | 0.021 | 0.870 | 0.054 |
| Fine | 256,811 | 0.0068 | 0.021 | 0.870 | 0.054 |
| Experiment [27] | - | 0 | 0.024 | 1.03 | 0.06 |

## 4. Results and discussion

### 4.1 Lift and drag coefficient

Lift and drag coefficients for the four RANS models considered herein, namely k-$k_l$-ω, SA, k-ω SST and non-linear k-ω SST, are plotted for 0, 5, 10 and 12 degrees AOA in Figure 4(a) and 4(b). In Figure 4(a), one observes that k-$k_l$-ω, SA, and k-ω SST follow a linear trend whereas only non-linear k-ω SST model is able to correctly predict the linear as well as non-linear trend both qualitatively and quantitatively. In the linear regime, both linear and non-linear k-ω SST model have correct predictions of $C_L$ but k-$k_l$-ω and SA models under-predict the lift coefficient. At 12 degrees AOA, linear growth trend of former three models results in over-prediction of $C_L$ for SST model and correct prediction of $C_L$ for k- $k_l$ -ω and SA model. But only, non-linear k-ω SST model seems to leave the linear trend and correctly predict the lift coefficient at 12 degrees AOA.

In Figure 4(b), k-$k_l$-ω and NSST models show a very close prediction of the $C_D$ with respect to the experiment [27]. But, k-ω SST and SA models exhibit huge under-prediction due to absence of laminar separation bubble on the suction surface of the airfoil. It is quite interesting to note that the non-linear correction has significantly improved the prediction of laminar separation bubble, which can be further illustrated in the following subsections (4.2-4.3).

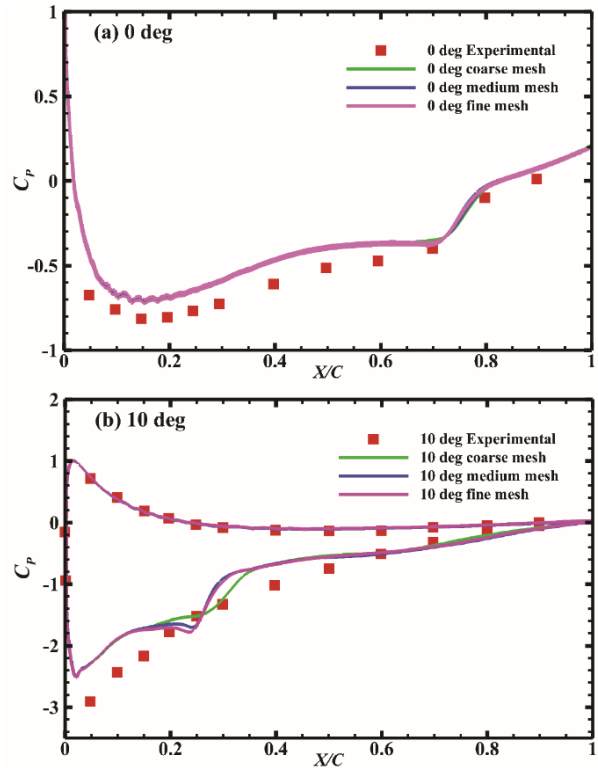

Fig. 3. Comparison of surface pressure coefficient obtained for k-$k_l$-ω model with experimental results in [27] among different grid resolution

### 4.2 Pressure coefficient

Figure 5(a)-5(d) shows variation of pressure coefficient along the direction of chord length at 0, 5, 10 and 12 degrees AOA for all the RANS models. At all the angles of attack, SA and k-ω SST models fail to predict the presence of laminar separation bubble at the suction side of the airfoil. But, pressure coefficient predicted by k-$k_l$-ω and non-linear k-ω SST

models closely follow the experimental data [27] which is also reflected in the prediction of laminar separation bubble. The under-performance of SA and k-ω SST model can be attributed to the fact that these models have been developed for fully turbulent flows, hence they are not expected to predict flow transition. However, the k- $k_l$ -ω model has been designed specifically for the prediction of transitional flows which results in an excellent prediction of laminar separation bubble as evident from pressure coefficient plots. It is a well-known fact that the flow experiencing separation induced transition in laminar separation bubble region is associated with high anisotropy in Reynolds stress. Hence, in a 2D, URANS simulation, it becomes very important to correctly model all component of Reynolds stress to obtain an accurate account of mean flow quantities. Interestingly, non-linear corrections added to the k-ω SST model, has enabled the prediction of pressure coefficient in the transition region. This has made the wall pressure coefficients comparable (or better in case of 0 and 10 degree angle of attack) to k-$k_l$-ω model.

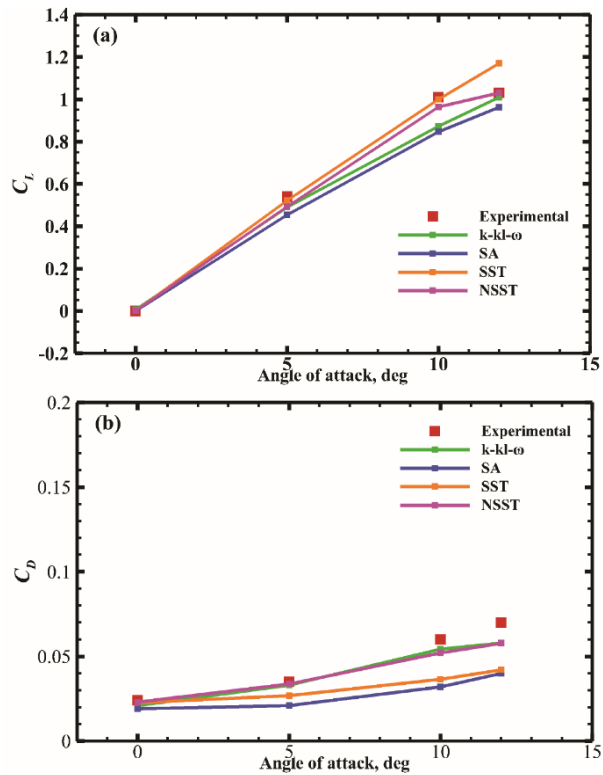

Fig. 4. Variation of lift and drag coefficient for NACA 0021 airfoil with AOA. Experimental results are taken from ref. [27].

As seen in Figure (6(a)-6(c)), while, k-ω SST and SA model are under-predicting all the components of Reynolds stress, non-linear k-ω SST model has very close prediction of stream-wise($< u'u' >$), wall normal ($< v'v' >$) and Reynolds shear stress ($< u'v' >$) components of turbulent stress with the k-$k_l$-ω model. This results in a very accurate prediction of mean flow quantities and consequently a good match of lift and drag coefficients are obtained.

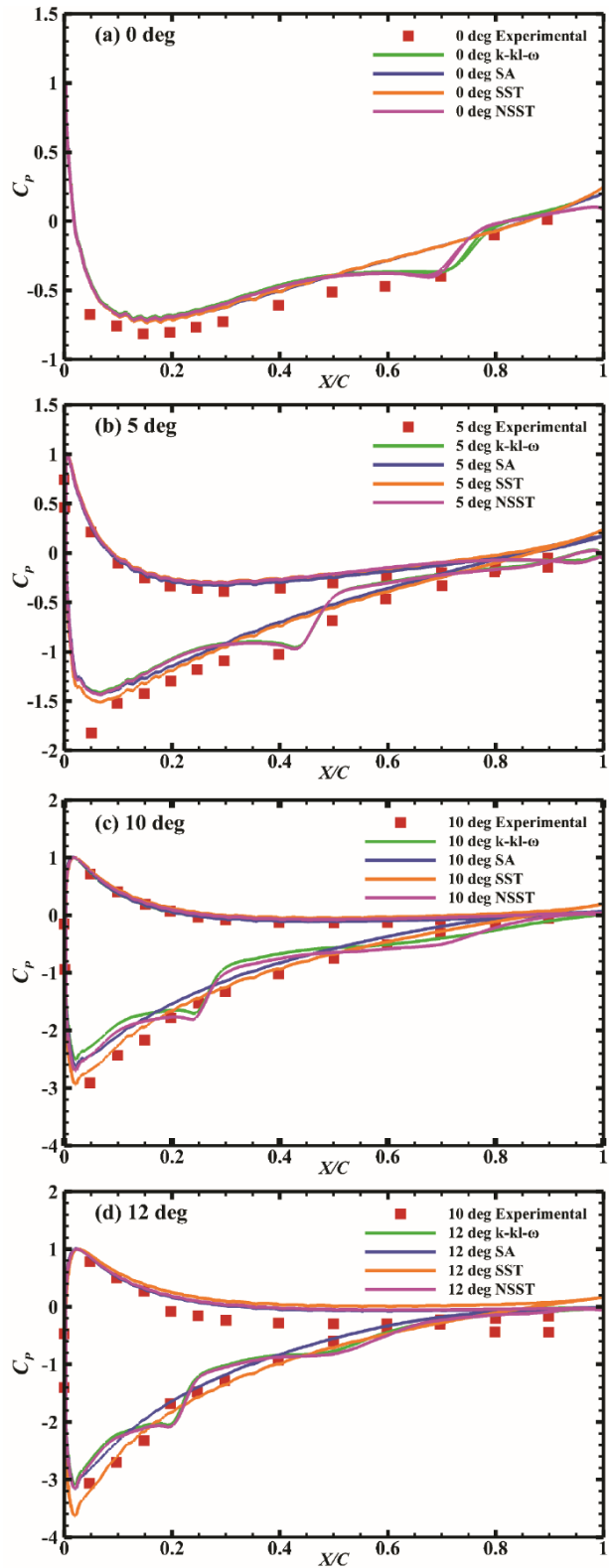

Fig. 5. Pressure Coefficient distribution of NACA 0021 at 0, 5, 10 and 12 degrees AOA. Experimental results are taken from [27].

Forces arising due to pressure distribution on a thin symmetric airfoil surface contribute to both lift and drag coefficients as follows:

$$C_{L_p} = \cos(\alpha) \int_0^C \left( C_{p_l} - C_{p_u} \right) dx \qquad (18)$$

$$C_{D_p} = \sin(\alpha) \int_0^C \left( C_{p_l} - C_{p_u} \right) dx \qquad (19)$$

where "$\alpha$" is the AOA, $C_{L_p}$ and $C_{D_p}$ represents lift and drag coefficient due to pressure, respectively while "l" and "u" stands for lower and upper surfaces, respectively.

Here $\int_0^C \left( C_{p_l} - C_{p_u} \right) dx$ represents the area enclosed by the $C_p$ curve.

Hence, correct prediction of laminar separation bubble length and associated decrease in pressure within LSB is an important parameter to characterize both lift and drag coefficients at high angles of attack. At 0 degree AOA, the area under $C_p$ curve is zero, which indicates the correct prediction of $C_{L_p}$ and $C_{D_p}$ for all the models.

### 4.3 Skin friction coefficient

Figure 7(a)-7(d) shows a variation of skin friction coefficient along the direction of chord length at 0, 5, 10 and 12 degrees AOA for all the RANS models. The difficulty in the prediction of correct wall shear stress using RANS model is well-known to the research community and the same is observed in case of k-ω SST and Spalart-Allmaras model. SA model shows a better prediction of wall shear stress than k-ω SST model; however, it is unable to predict the transitional flow regime, in turn LSB correctly while compared with conventional transitional model (k-$k_l$-ω). The available literature provides no experimental wall shear stress data except the separation and the reattachment points of laminar separation bubble as tabulated in Table 2. Also, k-$k_l$-ω seams to predict $C_f$ most accurately, hence it can be used as a benchmark for further analysis. For external flows (under turbulent condition), it is quite well known that SA model predictions are better compared to linear k-ω SST model and that's why it has been accepted widely for design purposes. But, the addition of non-linear Reynolds stress correction term to k-ω SST model has significantly improved the models capability for wall shear stress prediction which results in comparable or better (in case of 0 degree AOA) prediction with respect to the k-$k_l$-ω model. From Table 2, it can be seen that both k-$k_l$-ω and non-linear SST model are able to accurately predict the onset and reattachment point of laminar separation bubble except in the case of 0 degree AOA where k-$k_l$-ω shows over prediction in length of LSB compared to NSST model.

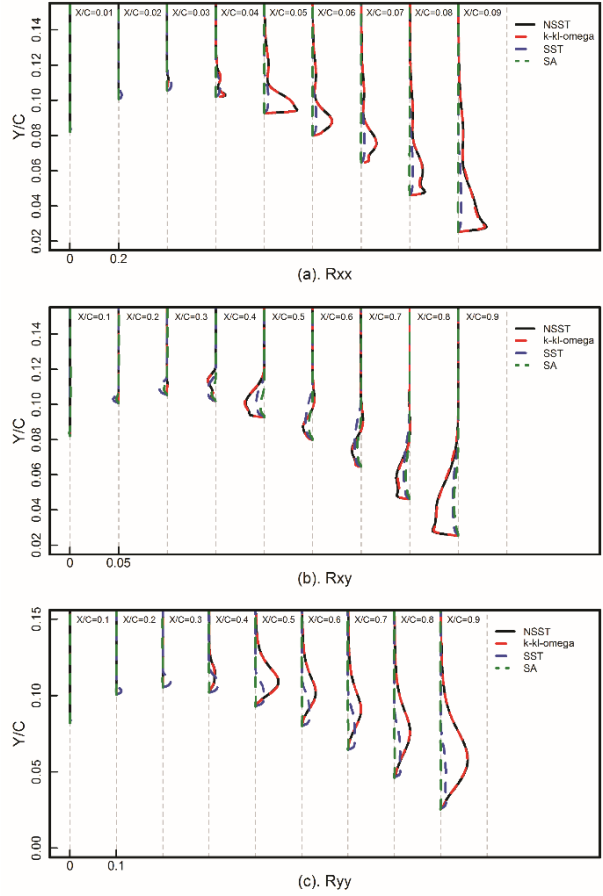

Fig. 6. (a) $<u'u'>$, (b) $<u'v'>$ and (c) $<v'v'>$ of NACA 0021 airfoil at 5 degrees AOA

Table2: Location and size of laminar separation bubble on the airfoil

| AOA | Experimental [27] | $k$-$kl$-$\omega$ | NSST |
|---|---|---|---|
| 0 | $0.440 \leq x/c \leq 0.730$ (LBS size = 0.290C) | $0.421 \leq x/c \leq 0.821$ (LBS size = 0.400C) | $0.422 \leq x/c \leq 0.760$ (LBS size = 0.338C) |
| 5 | $0.230 \leq x/c \leq 0.440$ (LBS size = 0.210C) | $0.229 \leq x/c \leq 0.501$ (LBS size = 0.272C) | $0.229 \leq x/c \leq 0.501$ (LBS size = 0.272C) |
| 8 | $0.100 \leq x/c \leq 0.250$ (LBS size = 0.150C) | $0.102 \leq x/c \leq 0.298$ (LBS size = 0.196C) | $0.102 \leq x/c \leq 0.293$ (LBS size = 0.191C) |
| 12 | $0.080 \leq x/c \leq 0.220$ (LBS size = 0.120C) | $0.0765 \leq x/c \leq 0.250$ (LBS size = 0.173C) | $0.0765 \leq x/c \leq 0.250$ (LBS size = 0.173C) |

Forces arising due to wall shear stress on a thin symmetric airfoil surface contribute to both lift and drag coefficients as follows:

$$C_{L_v} = -\sin(\alpha) \int_0^C \left( C_{f_l} + C_{f_u} \right) dx \qquad (20)$$

$$C_{D_v} = \cos(\alpha) \int_0^C \left( C_{f_l} + C_{f_u} \right) dx \qquad (21)$$

where "$\alpha$" is the AOA, $C_{L_v}$ and $C_{D_v}$ represents lift and drag coefficient due to wall shear stress, respectively and "l" and "u" stands for lower and upper surfaces, respectively.

Here $\int_0^C \left( C_{p_l} + C_{p_u} \right) dx$ represents twice of the average of $C_f$ curve.

Hence, correct prediction of laminar separation bubble length and a decrease in skin-friction coefficient inside LSB is also important in order to correctly predict both lift and drag coefficients at high angles of attack. But wall shear stress coefficient is an order of magnitude less than pressure coefficient at the wall, so the contribution of $C_f$ in the total lift and drag coefficient is very small. Hence, the differences seen in Figure 4(a) and 4(b) are mostly due to variation in the prediction of $C_p$ curve by different RANS models.

### 4.4 Laminar separation bubble

Prediction of laminar separation bubble is a challenge for most of the RANS simulation because RANS models are developed with the assumption of attached and fully turbulent boundary layer flow. Since, laminar separation bubble observes separation induced transition which demands resolution of high anisotropy in the bubble region for RANS computation, a transitional RANS model (k-$k_l$-ω) or an anisotropy resolving RANS model (non-linear k-ω SST) is necessary for an accurate prediction. In Figure 8 and 9, mean stream-wise velocity ($U_x$), stream-wise Reynolds stress ($< u'u' >$), wall-normal Reynolds stress ($< v'v' >$) and Reynolds shear stress ($< u'v' >$) has been plotted for all the RANS models at 5 and 10 degrees AOA. No separation bubble is seen at either of the AOAs for k-ω SST and Spalart-Allmaras models. But, exactly same position and size of LSB is observed for non-linear k-ω SST and k-$k_l$-ω model as shown in Figure 7. This difference is due to poor performance of linear models in prediction of anisotropy in Reynolds stress components. The k-ω SST and SA model show a very small magnitude of Reynolds stresses but a high magnitude of Reynolds stresses are observed in the case of k-$k_l$-ω and non-linear k-ω SST model which is the main reason for correct prediction of LSB. Also, the distribution of Reynolds stresses at different locations along the chord length presented in Figure 6(a)-6(c) corroborates the discussion in this section and validates the use of non-linear corrections in k-ω SST model for better prediction of laminar separation bubble.

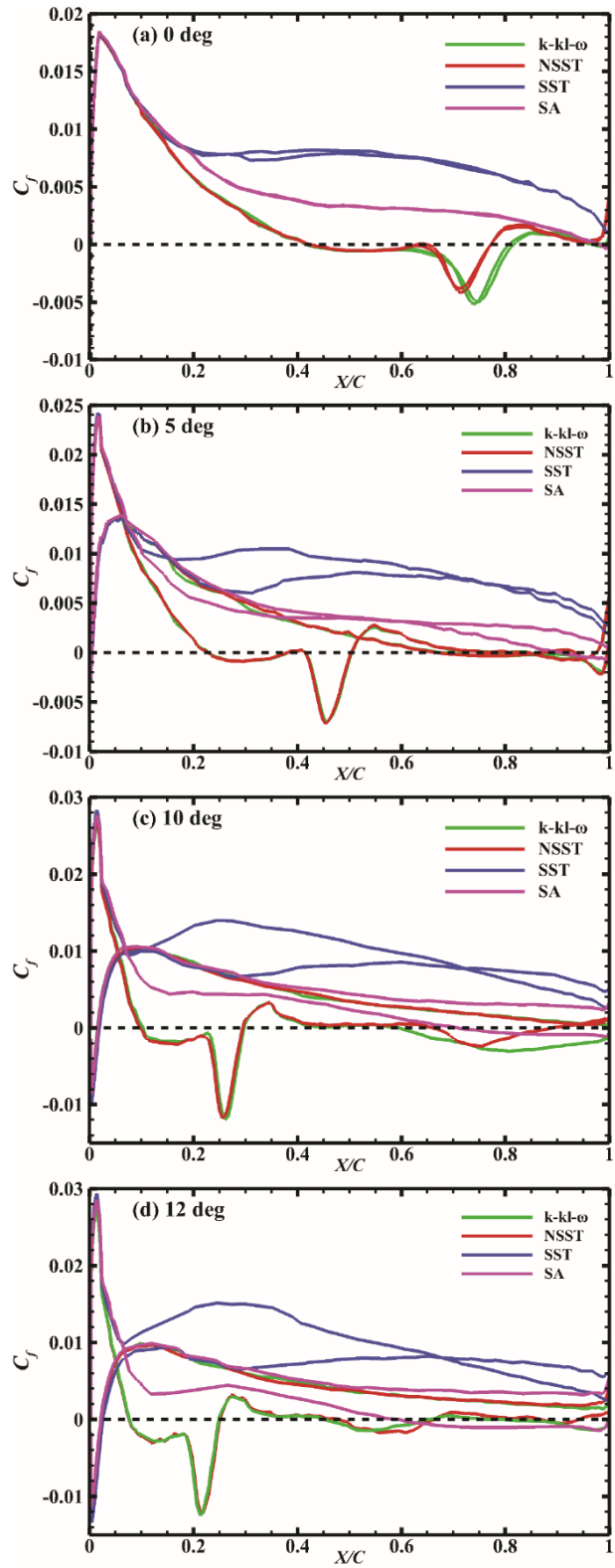

Fig. 7. Skin friction coefficient for NACA 0021 airfoil at selected angle of attack

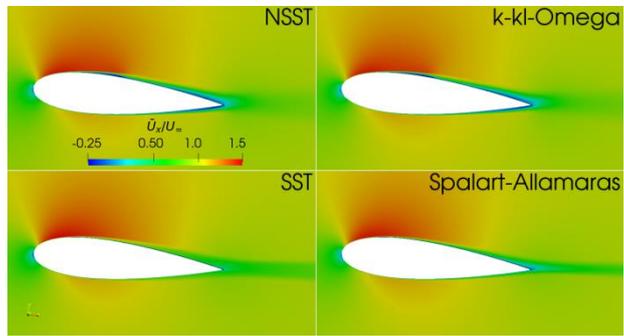

(a)  $\bar{U}_x/U_\infty$

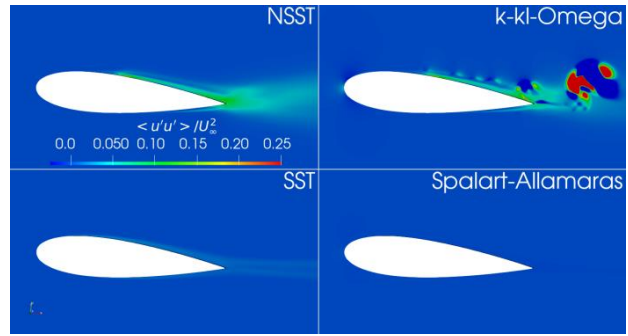

(b)  $< u'u' > / U_\infty^2$

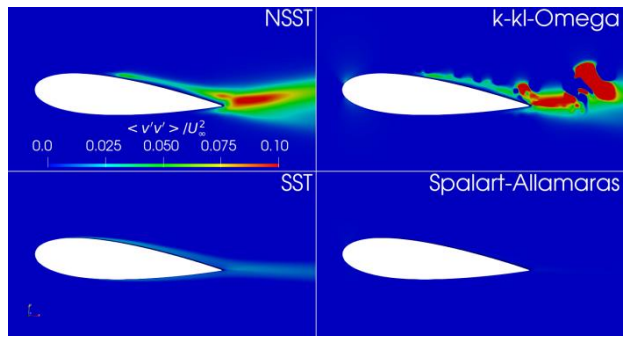

(c)  $< v'v' > / U_\infty^2$

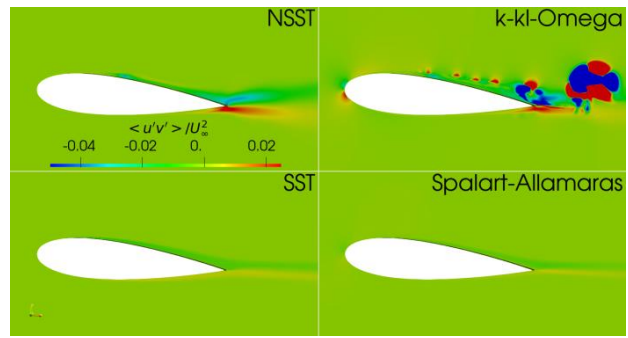

(d)  $< u'v' > / U_\infty^2$

Fig. 8. Stream-wise mean velocity, stream-wise and wall-normal Reynolds normal stress and Reynolds shear stress for all RANS models at 5 degrees AOA

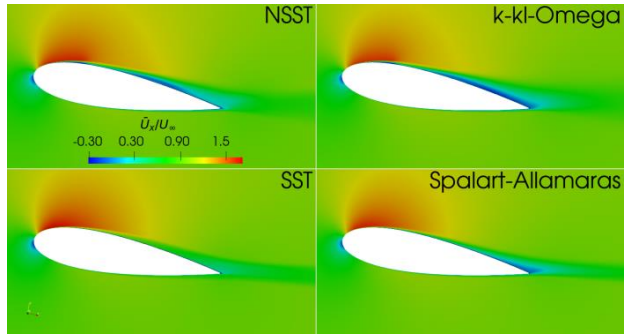

(a)  $\bar{U}_x/U_\infty$

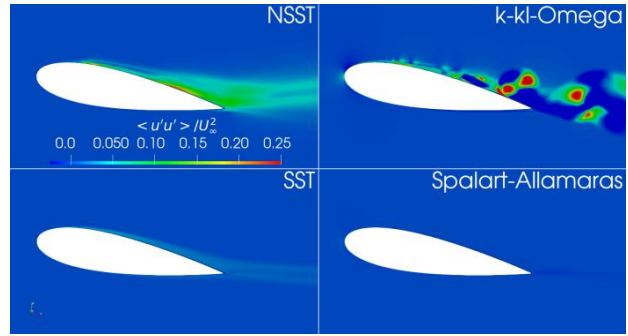

(b)  $< u'u' >/ U_\infty^2$

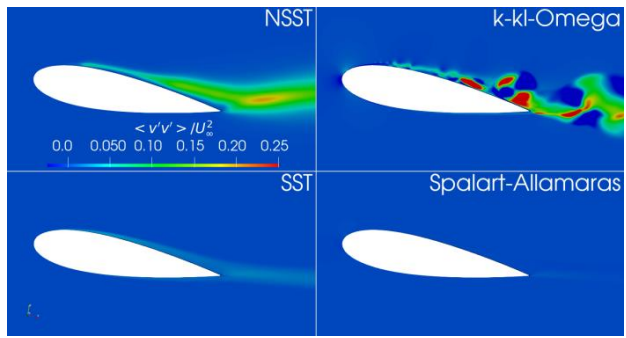

(d)  $< v'v' >/ U_\infty^2$

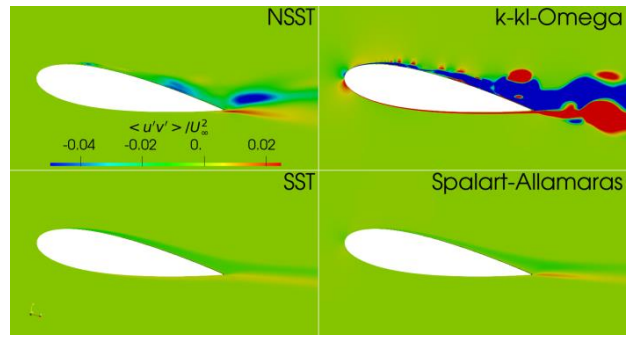

(d)  $< u'v' >/ U_\infty^2$

Fig. 9. Stream-wise mean velocity, stream-wise and wall-normal Reynolds normal stress and Reynolds shear stress for all RANS models at 10 degrees AOA

### 4.5 Computation time

A computation time study has also been performed to verify the economical efficiency gained by different turbulence models. All models have been run for 0.1 million iterations with 30 processors after the flow has been fully developed to avoid any computational bottleneck in initial transient flow. These computational bottlenecks arise due to uniform initialization of the flow field in the internal domain as initial condition. Computational time for this test has been tabulated in Table 3.

It can be observed from Table 3, that SA, k-ω SST, non-linear k-ω SST and k-$k_l$-ω show increasing computation time due to increase in a number of transport equation in turbulence models. But, non-linear k-ω SST model shows very small increase in computation time compared to it's linear counter-part. However, it can provide results with accuracy comparable to k-$k_l$-ω model.

Table3: Computational time

| Model | Time (in sec) for Model 0.1 million iterations with 30 processors |
|---|---|
| SA | 17457 |
| k-ω SST | 21575 |
| Non-linear k-ω SST | 21721 |
| k-$k_l$-ω | 31757 |

### 4.5 Validation of non-linear SST turbulence model for different airfoil cases

Separation induced transition is very crucial part in prediction of laminar separation bubble in airfoils at low Reynolds number. This exercise has been carried out in order to find the separation induced transition prediction capability of the non-linear SST model for different airfoils. To validate the prediction of separation induced transition with non-linear SST model, two more airfoils have been selected, namely NACA 0012 at 6° AOA and NACA 65021 at 8° AOA.

**NACA 0012:** A simulation has been performed over a thin airfoil NACA 0012 with linear k-ω SST and non-linear SST turbulence models at Re = 100,000 (based on chord length) for 6° angle of attack. The grid independence has been performed for NACA0012 at 6° angle of attack and a two dimensional structured grid with a total cells 141,000 is selected for further numerical simulation. Near wall grid is resolved upto $y^+ = 1$ in order to resolve the boundary layer correctly. The results are compared with the published literature [28] as shown in Figure 10(a) and 10(b). It can be observed that NSST model has a very good capability to predict separation induced transition in case of NACA 0012 airfoil. It can be seen that the linear turbulence model (SST) has good prediction of surface pressure coefficient all over the airfoil sections except in the separation induced transition region. The comparison of laminar separation bubble provided in Table 4 shows that prediction of laminar separation bubble is closely reproduced with non-linear k-ω SST model as compared to results in ref. [28].

Table4: Comparison of laminar separation bubble length for NACA 0012 with ref. [28]

| Laminar separation Bubble Position | NSST | Shau et al. [28] |
|---|---|---|
| Laminar separation point (x/c) | 0.064 | 0.06 |
| Turbulent reattachment point (x/c) | 0.350 | 0.285 |
| Laminar separation bubble Length (l/c) | 0.286 | 0.225 |

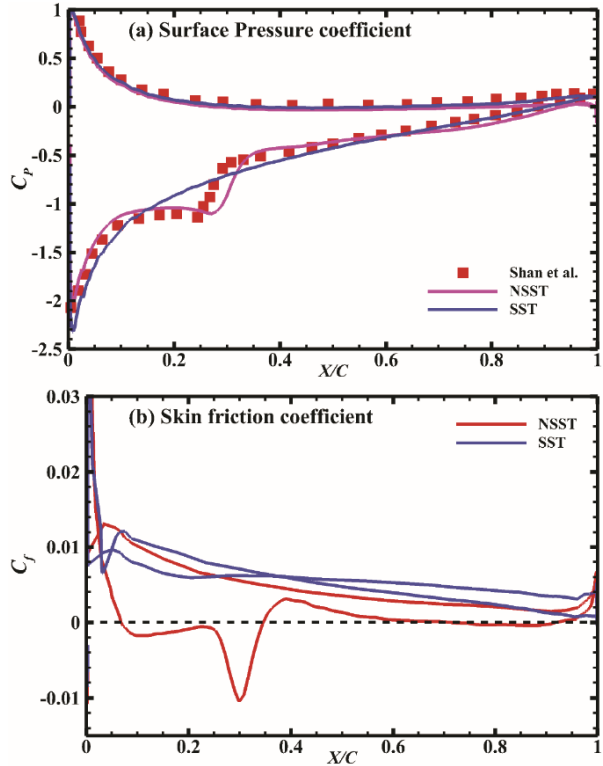

Fig. 10. Comparison of surface pressure coefficient and skin friction coefficient obtained for NACA 0012 using NSST model with published results in ref. [28] at 6° AOA

**NACA 65021:** A second test for laminar separation bubble prediction capability of non-linear k-ω SST model is chosen with NACA 65021 airfoil, which finds its application in wind turbine [19]. The grid independence has been performed for NACA65021 at 8° angle of attack and a two dimensional structured grid with a total cells 181,461 is selected for further numerical simulation. Near wall grid is resolved upto $y^+ = 1$ in order to resolve the boundary layer correctly. The simulated results of non-linear SST turbulence model provide excellent predictions of the pressure coefficient as compared to experimental data [27] as shown in Figure 11(a) and 11(b) and the linear k-ω SST model failed to predict laminar separation bubble. A quantitative comparison of laminar separation bubble generated on pressure surface of the airfoil is provided in Table 5, which shows very good match with the experimental results [27].

Table5: Comparison of laminar separation bubble length for NACA 65021 with ref. [27]

| Laminar separation Bubble Position | NSST | Experiment [27] |
|---|---|---|
| Laminar separation point (x/c) | 0.0765 | 0.080 |
| Turbulent reattachment point (x/c) | 0.250 | 0.22 |
| Laminar separation bubble Length (l/c) | 0.173 | 0.12 |

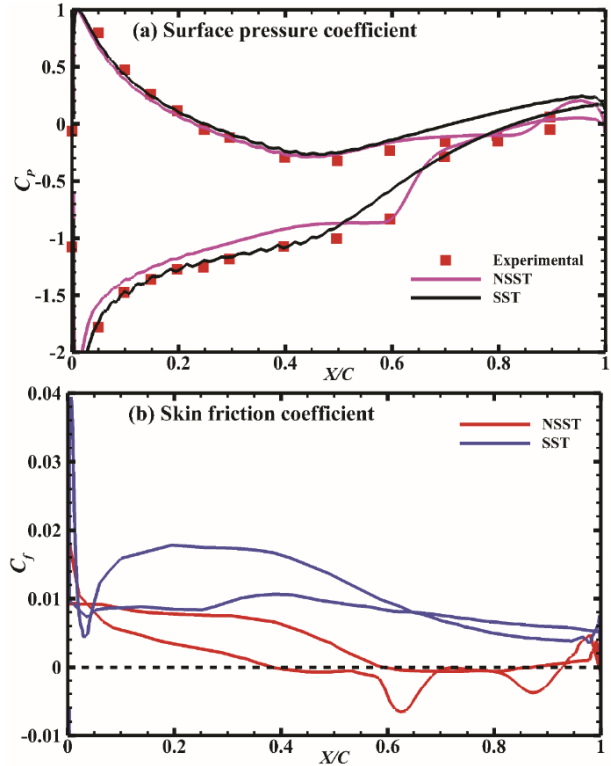

Fig. 11. Comparison of surface pressure coefficient and skin friction coefficient obtained for NACA 65021 using NSST model with published results in ref. [27] at 8° AOA

## 5. Conclusions

A two-dimensional numerical study has been carried out over NACA 0021 airfoil with four RANS models namely Spalart-Allmaras model, k-$k_l$-ω and k-ω SST model with and without non-linear corrections at Re = 120,000 for various angles of attack. Spalart-Allmaras model and linear k-ω SST models are unable to predict laminar separation bubble due to model constraints arising from the fully turbulent flow assumptions in its formulations. The k-$k_l$-ω model has appropriate production and dissipation terms related to natural and bypass transitions which enable it to correctly predict the laminar separation bubble in this case. The improvement suggested in this paper with the inclusion of non-linear corrections in k-ω SST model has been justified in the results and discussion section by accurate prediction of lift and drag coefficients, surface pressure coefficient, skin-friction coefficient and components of Reynolds stress tensor. The inclusion of non-linear corrections in k-ω SST model introduces the prediction capability for separation induced transition and laminar separation bubble.

Based on the testing in the current study, it is proposed that the use of non-linear corrections in k-ω SST model adds the capability to predict anisotropy in the Reynolds stress and consequently predict specific types of transitions such as one seen in laminar separation bubble.

## Acknowledgment

Simulations are carried out on the computers provided by the Indian Institute of Technology Kanpur (IITK) (www.iitk.ac.in/cc) and the manuscript preparation, as well as data analysis, has been carried out using the resources available at IITK. This support is gratefully acknowledged.

## Nomenclature

AOA : angle of attack
C : airfoil chord length
$C_L$ : lift coefficient
$C_D$ : drag coefficient
$C_P$ : pressure coefficient
k : turbulent kinetic energy
LBS : laminar separation bubble
$y^+$ : nondimensional wall distance
Re : Reynolds number
$\omega$ : specific turbulent dissipation rate
$\nu$ : molecular kinematic viscosity
$U_\infty$ : freestream velocity
RANS : Reynolds averaged Naviar Stokes

**Subscripts**
i,j : indices
$\infty$ : freestream condition